\documentclass[prd,twocolumn,amsmath,amssymb]{revtex4-2}

\usepackage{graphicx}	
\usepackage{subfigure}	
\usepackage[export]{adjustbox}

\newcommand\nn{\nonumber}
\newcommand\eq[1] {\begin{align} #1 \end{align}}

\newcommand{\br}[1]{\left( #1 \right)}
\newcommand{\brs}[1]{\left[ #1 \right]}

\newcommand{\brm}[1]{\left| #1 \right|}

\newcommand{\bra}[1]{\left< #1 \right|}
\newcommand{\ket}[1]{\left| #1 \right>}

\newcommand{\Sp}{\mbox{Sp}}

\renewcommand{\Re}{\mbox{Re}}

\newcommand{\vv}[1]{{\bf #1}}

\newcommand{\dd}[1]{{\hat #1}}        

\newcommand{\M} {{\cal M}} 						
\newcommand{\LQCD}{\Lambda_{\mbox{\tiny QCD}}}	

\newcommand{\GeV}{\mbox{GeV}}
\newcommand{\MeV}{\mbox{MeV}}

\newcommand{\eV} {\mbox{eV}}

\begin{document}

\title{The cross section of the process $e^+e^- \to \Lambda\bar{\Lambda}$ in the vicinity of charmonium $\psi(3770)$ including three-gluon and $D$-meson loop contributions}

\author{Yu.M.~Bystritskiy}
\email{bystr@theor.jinr.ru}
\affiliation{Joint Institute for Nuclear Research, 141980 Dubna, Moscow Region, Russia}

\author{A.~I.~Ahmadov}
\email{ahmadov@theor.jinr.ru}
\affiliation{Joint Institute for Nuclear Research, 141980 Dubna, Moscow Region, Russia}
\affiliation{Institute of Physics, Azerbaijan National Academy of Science, Baku, Azerbaijan}

\date{\today}

\begin{abstract}
The total cross section of the process $e^+e^- \to \Lambda\bar{\Lambda}$ is calculated within
the energy range
close to the mass of $\psi(3770)$ charmonium state. Two different contributions were considered:
the $D$-meson loop
and the three gluon charmonium annihilation one. Both of them contribute noticeably and
in sum fairly reproduce the data. Large relative phase for these contributions are generated
with respect to the pure electromagnetic mechanism.
As a by product the fit for the electromagnetic form factor of $\Lambda$-hyperon is obtained for
the large momentum transferred region.
\end{abstract}

\maketitle

\section{Introduction}

The bound state of a pair of charmed quarks are the one of the
most clear and simple system which allows one to study the details of
a confinement mechanism and to refine the conjectures of Quantum Chormodynamics (QCD).
During few recent decades these states are under intensive experimental and theoretical study
\cite{Asner:2008nq,Kussner:2020yfa,Nerling:2021bxo,Yuan:2021wpg}.
The special interest is focused to the electron--positron annihilation
processes with the production of different mesonic and baryonic final states
thus giving us the clear way to produce and to study the charmonium in pure $J^{PC}=1^{--}$ state.
The binary final states (i.e. with two particles finally produced) give the possibility
to further simplify the consideration of the processes with charmonium in the
intermediate state \cite{DM2:1987riy}.

Besides it was shown a long time ago that these processes are the excellent
way to measure the electromagnetic form factors of the particles produced
\cite{Cabibbo:1960zza, Cabibbo:1961sz}.
There is enormous set of the measurements of the electromagnetic form factors
of the proton, for example, by the BaBar Collaboration \cite{BaBar:2013ves, BaBar:2013ukx}
or by the BES III Collaboration \cite{BESIII:2015axk}.
But the other baryons are also under study:
$\Lambda^0$ and $\Sigma^0$ baryons \cite{BaBar:2007fsu, Dobbs:2017hyd},
$\Sigma^\pm$ baryons \cite{BESIII:2020uqk},
$\Xi^0$, $\Xi^-$ and $\Omega^-$ baryons \cite{Dobbs:2014ifa}.
For a recent review of the situation on $\Lambda\bar{\Lambda}$ pair production see
a review \cite{Zhou:2022jwr}.
A lot of interest paid to the near threshold behaviour of these form factors
\cite{BESIII:2017hyw, BESIII:2020ktn} which demonstrates non-trivial enhancement effect.
Many theoretical ideas to explain this effect were proposed 
(see, for example, \cite{Haidenbauer:2006dm, Li:2021lvs}).

In this paper we want to consider the process of electron-positron annihilation
into a pair of $\Lambda$-hyperons which was recently precisely measured
in the vicinity of $\psi(3770)$ charmonium \cite{BESIII:2021ccp}.
The charmonium $\psi(3770)$ is the one of the intriguing states
which was studied by many collaborations (for example, by
KEDR-VEPP-4M \cite{Baldin:2008zz, Anashin:2011kq}, CLEO \cite{Ge:2008aa}
and more recently by BES III \cite{Ablikim:2008zz, Ablikim:2014jrz}).
In paper \cite{Ablikim:2014jrz} one can find the measurement of the cross section
of the process $e^+ e^- \to p\bar{p}$ with the specific care to the region near the 
mass of $\psi(3770)$ which demonstrates the dip instead of enhancing Breit--Wigner peak.
This is the manifestation of the large relative phase which is generated by the
intermediate charmonium state with respect to the pure electromagnetic background.
In papers \cite{Ahmadov:2013ova, Bystritskiy:2021frx} we showed that the source of this phase is
mostly attributed to the three gluon mechanism of the charmonium decay.
Here we consider the similar process $e^+ e^- \to \Lambda\bar{\Lambda}$
which also must have large relative phase coming from the same three gluon mechanism.
Since all the calculations was already done in \cite{Bystritskiy:2021frx} here
we just briefly recall the main formulae and put the main focus on the
details which differ in $\Lambda\bar{\Lambda}$ case.

The paper is organized in the following manner:
in Section~\ref{sec.Born} the total cross section of the process
$e^+e^- \to \Lambda\bar{\Lambda}$ in Born approximation is presented,
the electromagnetic form factors of $\Lambda$-hyperon are discussed;
Section~\ref{sec.PsiIntermediateState} shows how the charmonium in the intermediate state contributes
to the total cross section;
two subsequent Sections~\ref{sec.DMesonLoopMechanism} and \ref{sec.ThreeGluonMechanism}
shows the main formulae for the OZI-allowed mechanism with the $D$-meson loop and
for the OZI-violating three gluons mechanism.
Section~\ref{sec.Numerical} gives some numerical estimations and
comparison of our calculation with experimental data from BES~III \cite{BESIII:2021ccp};
Section~\ref{sec.Conclusion} concludes our results and proposes some possible
extension of this work in the future.

\section{Born approximation}
\label{sec.Born}

\begin{figure}
    \centering
    \includegraphics[width=0.25\textwidth]{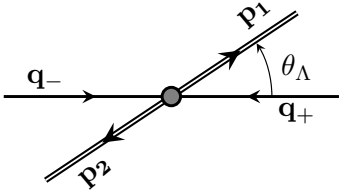}
    \caption{The definition of the scattering angle $\theta_\Lambda$ from (\ref{eq.Phi2}) in
    the center-of-mass reference frame.}
    \label{fig.MomentaPosition}
\end{figure}

We consider the process of electron-positron annihilation into a pair of $\Lambda$-baryons:
\eq{
    e^+(q_+)+e^-(q_-) \to \Lambda(p_1)+\bar{\Lambda}(p_2),
    \label{eq.Process}
}
where quantities in parenthesis are the 4-momenta of the corresponding particles.
The cross section for the process has the standard way:
\eq{
    d\sigma = \frac{1}{8 s} \sum_{\text{spins}} \brm{\M}^2 \, d\Phi_2,
    \label{eq.CrossSectionGeneralForm}
}
where the summation of the amplitude square $\brm{\M}^2$ runs over all possible initial and final particles spin states.
We systematically neglect the mass of the electron $m_e$ in this paper.
The phase volume of final particles $d\Phi_2$ has the form:
\eq{
    d\Phi_2 &=
    \frac{1}{\br{2\pi}^2} \delta\br{q_++q_- - p_1 - p_2} \frac{d\vv{p_1}}{2E_1}\frac{d\vv{p_2}}{2E_2}
    =
    \nn\\
    &=
    \frac{\brm{\vv{p}}}{2^4 \pi^2 \sqrt{s}} \, d\Omega_\Lambda
    =
    \frac{\beta}{2^4\pi} \, d\cos\theta_\Lambda,
    \label{eq.Phi2}
}
where $d\Omega_\Lambda=d\phi_\Lambda \, d\cos\theta_\Lambda$ and
$\phi_\Lambda$ and $\theta_\Lambda$ are the azimuthal and the polar angles of the final $\Lambda$-baryon momentum
in the center-of-mass reference frame (center of mass system, c.m.s),
i.e. $\theta_\Lambda$ is the angle between 3-momenta of the initial electron $\vv{q_-}$ and the final $\Lambda$-baryon
$\vv{p_1}$ (see Fig.~\ref{fig.MomentaPosition})
and $\brm{\vv{p}} \equiv \brm{\vv{p_1}} = \brm{\vv{p_2}} = \sqrt{s} \beta / 2$ is
the modulus of 3-momenta of the final $\Lambda$ or $\bar{\Lambda}$.
Here the quantity $\beta = \sqrt{1-4M_\Lambda^2/s}$ is the final $\Lambda$-baryon velocity,
with $M_\Lambda$ being the mass of $\Lambda$-baryon.

\begin{figure}
	\centering
    \subfigure[]{\includegraphics[width=0.27\textwidth]{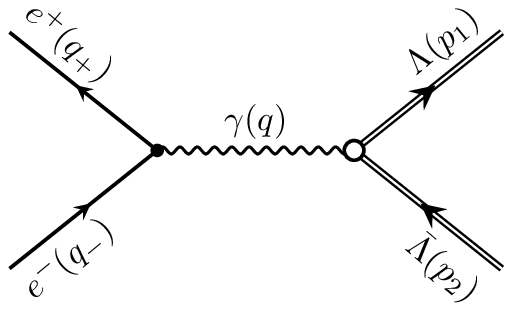}\label{fig.BornDiagram}}
	\hspace{0.05\textwidth}
    \subfigure[]{\includegraphics[width=0.27\textwidth]{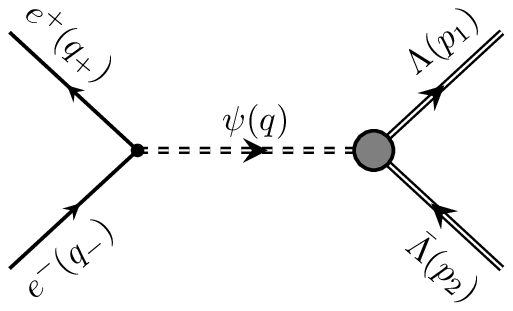}\label{fig.PsiDiagram}}
    \caption{
    	Feynman diagrams of the process $e^+e^- \to \Lambda \bar{\Lambda}$ in Born approximation (a)
    	and with the charmonium $\psi(3770)$ intermediate state (b).
    }
    \label{fig.TwoMechanisms}
\end{figure}

In Born approximation (see Fig.~\ref{fig.BornDiagram}) the amplitude $\M=\M_B$ in
(\ref{eq.CrossSectionGeneralForm}) describes the electron--positron pair annihilation into
virtual photon, which then produces $\Lambda\bar{\Lambda}$ pair. The amplitude $\M_B$ corresponding to this
process has the form:
\eq{
    \M_B = \frac{1}{s} J^{e\bar{e}\to\gamma}_\mu(q) \, J_{\gamma\to \Lambda\bar{\Lambda}}^\mu(q),
    \label{eq.BornAmplitude}
}
where $s = q^2 = \br{q_++q_-}^2 = \br{p_1+p_2}^2$
is the total invariant mass squared of the lepton pair ($q$ is the momentum of the
intermediate photon).
The quantities $J^{e\bar{e}\to\gamma}_\mu$ and $J^{\gamma\to \Lambda\bar{\Lambda}}_\mu$
are electromagnetic currents:
\eq{
    J^{e\bar{e}\to\gamma}_\mu(q) &= -e \brs{\bar{v}(q_+) \gamma_\mu u(q_-)},
    \label{eq.CurrentEEGamma}\\
    J^{\gamma\to \Lambda\bar{\Lambda}}_\mu(q) &= e \brs{\bar{u}(p_1) \Gamma_\mu(q) v(p_2)},
    \label{eq.CurrentLLGamma}
}
where $e$ is the modulus of electron charge $e = \sqrt{4\pi\alpha}$ with $\alpha$ being
the fine structure constant \cite{Zyla:2020zbs}. In general the vertex
of the photon with the $\Lambda$-baryon has the form:
\eq{
	\Gamma_\mu(q) &= F_1(q^2) \gamma_\mu - \frac{F_2(q^2)}{4M_\Lambda} \br{\gamma_\mu \dd{q} - \dd{q} \gamma_\mu}.
}
where we use the notation $\dd{a} \equiv a_\mu \gamma^\mu$.
Here the functions $F_{1,2}(q^2)$ are the $\Lambda$-baryon
electromagnetic form factors normalized as $F_1(0)=0$ and $F_2(0)=\mu_\Lambda$,
where $\mu_\Lambda$ is the $\Lambda$-baryon anomalous magnetic moment.

It was shown in \cite{Tomasi-Gustafsson:2020vae} that the non-trivial structure of the baryon starts
to manifest itself even at relatively low $q^2$ and thus one must take into account
these effects of the structure.
Since experimental data at the moment do not allow one to extract the electric $G_E$
and the magnetic $G_M$ form factors of a baryons separately we utilize the
assumption that $\brm{G_E} = \brm{G_M}$, i.e. $F_2(q^2)=0$.
Then the total cross section in Born approximation has the form:
\eq{
    \sigma_B(s)=\frac{2\pi\alpha^2}{3s} \beta \br{3-\beta^2} \brm{F_1\br{s}}^2.
    \label{eq.TotalCrossSectionBorn}
}
The form factor $F_1$ is chosen to have pQCD inspired form \cite{Lepage:1979za,Lepage:1980fj},
which takes into account the running of the QCD coupling constant $\alpha_s$:
\eq{
	F_1(s) = \frac{C}{s^2 \log^2\br{s/\LQCD^2}},
	\label{eq.Formfactor}
}
where $\LQCD$ is the QCD scale and the constant $C$ should be fitted on the experimental data
for baryon--antibaryon production in a wide energy range.

For proton in our energy region this fit was done in \cite{Ablikim:2014jrz}
giving $C = (62.6 \pm 4.1)~\GeV^4$ (using $\LQCD = 300~\MeV$).
This fit qualitatively agrees with the more recent result of paper \cite{Bianconi:2015owa} where
this constant was fitted to be equal to $C = 72~\GeV^4$ with $\LQCD = 520~\MeV$.

\begin{figure}
	\centering
	\includegraphics[width=0.48\textwidth]{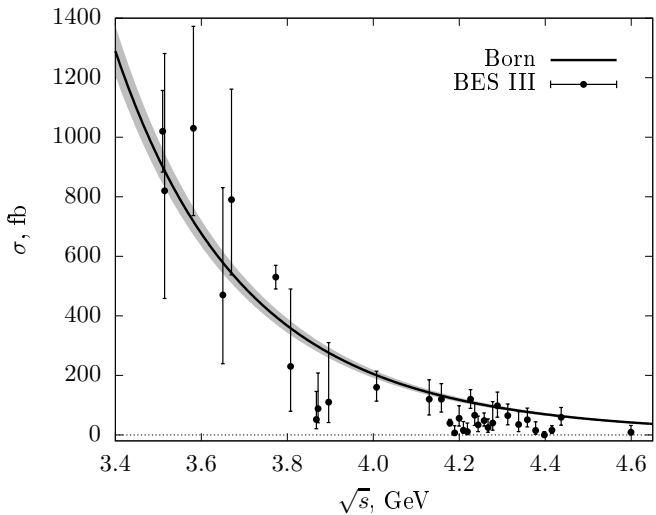}
    \caption{The total cross section for the process $e^+ e^-\to \Lambda \bar{\Lambda}$.
    Black line is the cross section in Born approximation (\ref{eq.TotalCrossSectionBorn}).
    The curve errors origin from the form factor constant (\ref{eq.C}) fitting errors.}
    \label{fig.WideRange}
\end{figure}

In our case of $\Lambda\bar{\Lambda}$ pair production we fix this constant
using the whole range of BES-III measurement \cite{BESIII:2021ccp} presented
in Fig.~\ref{fig.WideRange}.
Fitting the Born cross section from (\ref{eq.TotalCrossSectionBorn}) with respect to this
data gives us the following parameter:
\eq{
	C = \br{43.1 \pm 1.4}~\GeV^4,
    \label{eq.C}
}
which we use further for $\Lambda$-baryon electromagnetic form factor (\ref{eq.Formfactor}).
We note that this expression for the form factor $F_1$ with
the constant $C$ from (\ref{eq.C}) works for relatively large momentum transfered $q^2$.
It does not pretend to work near threshold, since there are many delicate
features playing important role there, such as Coulomb-like
enhancement factor \cite{Haidenbauer:2014kja,Amoroso:2021slc} or
the manifestation of wavy nature of baryon stabilization after its
emerging from the vacuum \cite{Tomasi-Gustafsson:2020vae}.

\section{The quarkonium $\psi(3770)$ intermediate state}
\label{sec.PsiIntermediateState}

The main task of our work is to describe the effect of the charmonium resonance $\psi(3770)$
excitation in the process (\ref{eq.Process}).
As one can see in Fig.~\ref{fig.WideRange} the experimental point for the cross section
at $\sqrt{s} = M_\psi$ (where $M_\psi$ is the $\psi(3770)$ mass)
flies higher the Born cross section curve.
In \cite{BESIII:2021ccp} there is a fit of this point with the use of some
phenomenologically inspired formula (see eq. (3) in \cite{BESIII:2021ccp}).
Further we develop the model (based on our previous calculations
\cite{Ahmadov:2013ova, Bystritskiy:2021frx}) which reveals the underlying mechanism for this point
to be upstairs.

In the region of charmonium resonance $\psi(3770)$ excitation one should take into account
the additional contribution to the amplitude:
\eq{
    \M=\M_B+\M_\psi,
}
where $\M_\psi$ takes into account the mechanism with charmonium $\psi(3770)$
in the intermediate state (see Fig.~\ref{fig.PsiDiagram}) which is enhanced by the
Breit-Wigner factor:
\eq{
    \M_\psi = \frac{g^{\mu\nu} - q^\mu q^\nu/M_\psi^2}{s-M_\psi^2+i M_\psi\,\Gamma_\psi} J^{e\bar{e}\to\psi}_\mu(q)
     J^{\psi\to \Lambda\bar{\Lambda}}_\nu (q),
    \label{eq.Mpsi}
}
where $\Gamma_\psi$ is the total decay width of $\psi(3770)$ resonance and
$J^{e\bar{e}\to\psi}_\mu$ and $J^{\psi\to \Lambda\bar{\Lambda}}_\mu$ are the currents which describe the transition
of lepton pair into $\psi(3770)$ resonance and the transition of the $\psi(3770)$ resonance into
$\Lambda\bar{\Lambda}$ pair correspondingly.
Following \cite{Ahmadov:2013ova} we assume that $J^{e\bar{e}\to\psi}_\mu$ has the same structure as
$J^{e\bar{e}\to\gamma}_\mu$ from (\ref{eq.CurrentEEGamma}), i.e.
\eq{
    J^{e\bar{e}\to\psi}_\mu(q) = g_e \, \brs{\bar{v}(q_+) \gamma_\mu u(q_-)},
}
where the constant $g_e = F_1^{\psi\to e\bar{e}}(M_\psi^2)$ is the value of the form factor of the
vertex $\psi \to e\bar{e}$ at the $\psi(3770)$ mass-shell
(here we follow the same approximation as in the Born case and
assume that $F_2^{\psi\to e\bar{e}}(M_\psi^2) = 0$).
We fix this constant $g_e$ via total decay width of $\psi\to e^+ e^-$ which is known to be equal to
$\Gamma_{\psi\to e^+e^-} = 261~\eV$ \cite{Zyla:2020zbs}:
\eq{
    g_e= \sqrt{\frac{12\pi\Gamma_{\psi\to e^+e^-}}{M_\psi}} = 1.6 \cdot 10^{-3}.
}
We neglect a possible imaginary part of the vertex $g_e$ since it
was shown in \cite{Kuraev:2013swa} that it is small, less then 10~\% of the real part.
We also note a mistake in our paper \cite{Bystritskiy:2021frx} where this constant is claimed to be
related with the $\psi \to p\bar{p}$ decay.

The amplitude $\M_\psi$ from (\ref{eq.Mpsi}) interferes with the Born one $\M_B$ giving
the standard interference contribution to the cross section:
\eq{
    d\sigma_{int} = \frac{1}{8s} \sum_{\text{spins}} 2\,\Re\brs{\M_B^+ \M_\psi} \, d\Phi_2,
}
which leads to the following form of the interference contribution to
the total cross section:
\eq{
	\sigma_{int}(s)
	=
	\Re\br{ \frac{S_i(s)}{s-M_\psi^2+i M_\psi\, \Gamma_\psi} },
	\label{eq.SigmaIntViaSi}
}
where $S_i(s)$ contains all the dynamics of the transformation of charmonium into
$\Lambda\bar{\Lambda}$ pair and has the following explicit form:
\eq{
	S_i(s)
    = \frac{e g_e \beta}{48 \pi s}
    	\int\limits_{-1}^1 d\cos\theta_\Lambda \sum_{s'}
    	\br{J_{\gamma\to \Lambda\bar{\Lambda}}^\alpha}^* J^{\psi\to \Lambda\bar{\Lambda}}_\alpha.
    \label{eq.Si}
}
The subscript index $i$ in the expression above denotes the type of mechanism of this transformation.
Since the mass of $\psi(3770)$ is higher than the threshold of $D$-meson pair
production it is natural to expect that the $D$-meson loop
will be the main mechanism in this reaction (see Fig.~\ref{fig.DD})
and we consider it below in Section~\ref{sec.DMesonLoopMechanism}.
However we need also to consider the OZI-violating three gluon mechanism (see Fig.~\ref{fig.3G})
which we briefly recall in Section~\ref{sec.ThreeGluonMechanism}.

Having the interference contribution (\ref{eq.SigmaIntViaSi}) with the total
relative phase between Born amplitude $\M_B$ and the charmonium contribution one $\M_\psi$
we can restore the total cross section using the procedure
described in \cite{Bystritskiy:2021frx} (see eqs.~(15) and (16) there).

\section{D-meson loop mechanism}
\label{sec.DMesonLoopMechanism}

\begin{figure}
	\centering
    \includegraphics[width=0.4\textwidth]{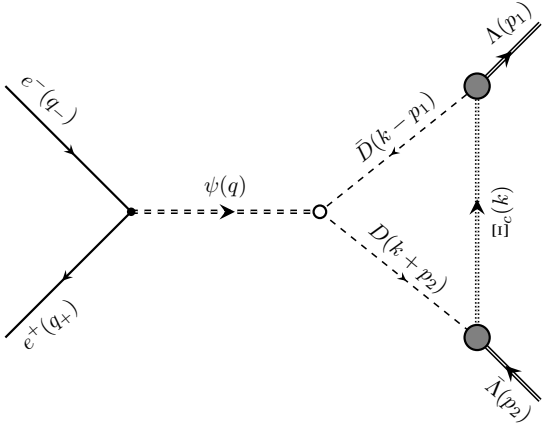}
    \caption{$D$-meson loop mechanism.}
    \label{fig.DD}
\end{figure}

Here we follow exactly to our previous calculations \cite{Ahmadov:2013ova, Bystritskiy:2021frx}
with only one systematic modification which is needed to describe
the $\Lambda\bar{\Lambda}$ pair final state instead of the proton--antiproton one.
The $D$-meson loop mechanism (presented in Fig.~\ref{fig.DD})
contribute to the interference of a charmonium state with the Born amplitude (see (\ref{eq.Si})) as:
\eq{
	S_D\br{s} &= \alpha_D\br{s} \, Z_D\br{s},
	\label{eq.SD}
}
where
\eq{
	&\alpha_D\br{s}
	=
	\frac{\alpha~g_e}{2^4 \, 3 \pi^2} \beta\,F_1(s),
    \nn\\
	&Z_D\br{s}
	=
	\frac{1}{s}
    \int\frac{dk}{i \pi^2}
    \times\nn\\
    &\times
    \frac{SpD(s,k^2)}
    {\br{k^2 - M_\Xi^2}\br{(k-p_1)^2 - M_D^2} \br{(k+p_2)^2 - M_D^2}}
    \times\nn\\
    &\quad\times G_{\psi D\bar{D}}(s,(k+p_2)^2,(k-p_1)^2)
    \times\nn\\
    &\quad\times G_{\Lambda D \Xi}\br{k^2,(k-p_1)^2} G_{\Lambda D \Xi}\br{k^2,(k+p_2)^2},
    \label{eq.ZD}
}
and $SpD(s,k^2)$ is the trace of $\gamma$-matrices over the baryon line:
\eq{
	&SpD(s,k^2)
	=
	\nn\\
	&=
    \Sp\brs{(\dd{p_1}+M_\Lambda) \gamma_5 (\dd{k}+M_\Xi) \gamma_5 (\dd{p_2}-M_\Lambda) (\dd{k} - M_\Lambda)}
    =\nn\\
    &=
    2\left(
    	\br{k^2}^2
    	+
    	k^2\br{
    		s - 2\br{M_D^2 + M_\Lambda M_\Xi}
    	}
    	-
    \right.
    \nn\\
    &\qquad\qquad\qquad\qquad\qquad\qquad\quad\left.\frac{}{}
    	-
    	s M_\Lambda M_\Xi
    	+
    	c_D
    \right),
    \label{eq.SpDExplicit}
    \\
    &c_D
    =
    M_D^4 + 2 M_\Lambda M_\Xi M_D^2 + 2 M_\Xi M_\Lambda^3 - M_\Lambda^4.
    \label{eq.cD}
}
The quantities $G_{\psi D\bar{D}}$ and $G_{D \Xi \Lambda}$ in (\ref{eq.ZD}) are
the form factors for the vertexes $\psi \to D\bar{D}$ and $D \to \Xi \Lambda$.

The details of the calculation of quantity $Z_D\br{s}$ can be found in \cite{Bystritskiy:2021frx},
but here we just recall that technically we calculate imaginary part of this quantity
and then restore the real part of it by using the dispersion relation with one subtraction
at $q^2=0$. First we need to mention that the subtraction constant here is also vanish since
the $\Lambda$-hyperon (which is the $uds$ quarks state) do not have open charm and thus
the vertex $\psi \to \Lambda \bar{\Lambda}$ at $q^2=0$ is zero.

Next, cutting the diagram by $D$-meson propagators we get the vertex
$\psi \to D\bar{D}$ with the only dependence over charmonium
virtuality $q^2 = s$ since $D$-meson legs are on-mass-shell:
\eq{
	G_{\psi D\bar{D}}\br{s,M_D^2,M_D^2}
	=
	g_{\psi D\bar{D}} \, \frac{M_\psi^2}{s} \, \frac{\log\br{M_\psi^2/\Lambda_D^2}}{\log\br{s/\Lambda_D^2}},
	\label{eq.PsiDDFormfactor}
}
where scale $\Lambda_D$ we fix on the characteristic value of the reaction $\Lambda_D = 2 M_D$
and the constant $g_{\psi D\bar{D}}$ is fixed by the $\psi \to D\bar{D}$
decay width:
\eq{
	g_{\psi D\bar{D}}
	&\equiv
	G_{\psi D\bar{D}}(M_\psi^2,M_D^2,M_D^2)
	=
	\nn\\
	&=
	4\,\sqrt{\frac{3\pi \, \Gamma_{\psi \to D\bar{D}}}{M_\psi \, \beta_D^3}}
	=
	18.4,
	\label{eq.gPsiDD}
}
where $\beta_D = \sqrt{1 - 4M_D^2/M_\psi^2}$ is the $D$-meson velocity in this decay.

Next we consider function $G_{\Lambda D \Xi}\br{k^2,p^2}$ from (\ref{eq.ZD}).
Again the only dependence left in the imaginary part of $Z_D$ is
the off-mass-shellness of $\Xi$ baryon in $t$-channel since $k^2 < 0$.
In \cite{Ahmadov:2013ova, Bystritskiy:2021frx} we used the following form of
$\Lambda D P$-vertex based on the results of \cite{Reinders:1984sr,Navarra:1998vi}:
\eq{
	G_{\Lambda D P}(k^2, M_D^2) = \frac{f_D \, g_{DN\Lambda}}{m_u + m_c},
	\qquad
	k^2 < 0,
}
where $f_D \approx 180~\MeV$ and
\eq{
	\frac{g_{DN\Lambda}}{\sqrt{4\pi}} = 1.9 \pm 0.6.
}
For quark masses the following values are used:
$m_u \approx 280~\MeV$ and  $m_c = 1.27~\GeV$ \cite{Zyla:2020zbs}.
The $SU(4)$ symmetry leads us to the same result for $G_{\Lambda D \Xi}$:
\eq{
	G_{\Lambda D \Xi}(k^2, M_D^2) = \frac{f_D \, g_{\Lambda D \Xi}}{m_u + m_c},
	\qquad
	k^2 < 0,
}
where
\eq{
	g_{\Lambda D \Xi} \approx g_{DN\Lambda} = 6.7 \pm 2.1.
	\label{eq.gLambdaDXi}
}

\section{Three gluon mechanism}
\label{sec.ThreeGluonMechanism}

\begin{figure}
	\centering
    \includegraphics[width=0.4\textwidth]{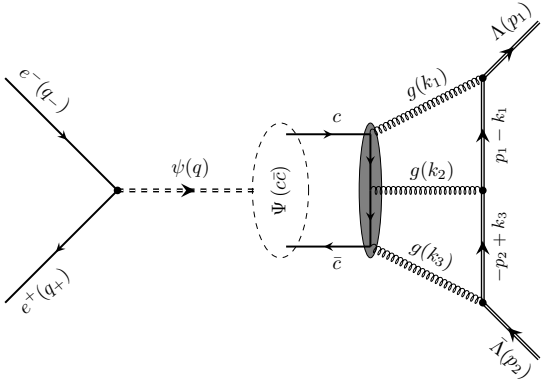}
    \caption{Three gluon mechanism.}
    \label{fig.3G}
\end{figure}

The three gluon mechanism first was considered in \cite{Ahmadov:2013ova} and
in \cite{Bystritskiy:2021frx} it was refined and some misprints and minor mistakes were corrected.
So here we just present the final formulae for its contribution to the
interference of a charmonium state with Born amplitude (see (\ref{eq.Si})):
\eq{
    S_{3g}\br{s}
    =
    \alpha_{3g} \br{s} \, Z_{3g}\br{s},
    \label{eq.S3g}
}
where
\eq{
    \alpha_{3g}\br{s} &= \frac{\alpha\, \alpha_s^3}{2^3 \, 3} g_e \, g_{col} \, \phi \, \beta \, F_1\br{s} \, G_\psi(s),
    \label{eq.alpha3g}
    \\
    Z_{3g}\br{s}
    &=
    \frac{4}{\pi^5 s}
    \int \frac{dk_1}{k_1^2}\frac{dk_2}{k_2^2}\frac{dk_3}{k_3^2}
    \times\nn\\
    &\times
    \frac{Sp3g ~ \delta\br{q-k_1-k_2-k_3}}{ (\br{p_1 - k_1}^2 - M_\Lambda^2) (\br{p_2 - k_3}^2 - M_\Lambda^2) },
    \label{eq.Z3g}
}
where the quantity $g_{col} = (1/4) \bra{\Lambda}  d^{ijk} ~ T^i T^j T^k \ket{\Lambda} = 15/2$ is the color factor
and $Sp3g$ is the product of traces over the $\Lambda$-hyperon and the $c$-quark lines:
\eq{
	Sp3g
	&=
    \Sp \brs{ \hat{Q}_{\alpha\beta\gamma} (\dd{p_c} + m_c) \gamma^\mu (\dd{p_{\bar{c}}} - m_c) }
    \times\nn\\
    &\times
   	\Sp\left[ (\dd{p_1}+M_\Lambda) \gamma^\alpha (\dd{p_1} - \dd{k_1} + M_\Lambda) \gamma^\beta \right.
	\times\nn\\
	&\qquad\times \left. (-\dd{p_2} + \dd{k_3} + M_\Lambda) \gamma^\gamma (\dd{p_2}-M_\Lambda) \gamma_\mu \right],
   	\nn
}
with
\eq{
    \hat{Q}_{\alpha\beta\gamma}
    &=
    \frac{\gamma_\gamma (-\dd{p_{\bar{c}}} + \dd{k_3} + m_c) \gamma_\beta (\dd{p_c} - \dd{k_1} + m_c) \gamma_\alpha}
    { ((p_{\bar{c}}-k_3)^2 - m_c^2) ((p_c-k_1)^2 - m_c^2) }
    +
    \nn\\
    &+ \brs{\text{gluon permutations}},
    \label{eq.DefinitionHatQ}
}
where the permutations over gluon vertices are performed in the gray block in Fig.~\ref{fig.3G}.

The quantity $\phi$ in (\ref{eq.alpha3g}) is related to the charmonium wave function
$\psi\br{\vv{r}}$ as:
\eq{
	\phi = \frac{\brm{\psi\br{\vv{r}=\vv{0}}}}{M_\psi^{3/2}} = \frac{\alpha_s^{3/2}}{3\sqrt{3\pi}},
	\label{eq.phi}
}
where $\alpha_s$ is the QCD coupling constant. We should note that this three gluon mechanism
is very sensitive to this quantity, since it
depends on its value in a rather high degree (see eqs.~(\ref{eq.alpha3g}) and (\ref{eq.phi})).
We use the value $\alpha_s(M_c) = 0.28$ which is expected
by the QCD evolution of $\alpha_s$ from the $b$-quark scale to the
$c$-quark scale, i.e. to $s \sim M_c^2$.
We note that this value differs from the one for the
$J/\psi$ charmonium which tends to feel much smaller value of parameter $\alpha_s(M_c) = 0.19$
\cite{Chiang:1993qi}.

The factor $G_\psi(s)$ in (\ref{eq.alpha3g}) is the form factor which describes
the mechanism of transition of three gluons (with total angular momentum equal to $1$)
into final $\Lambda\bar{\Lambda}$ pair. Following to \cite{Bystritskiy:2021frx}
we suggest that this mechanism has much in common with the proton--antiproton pair
production from the photon:
\eq{
	\brm{G_\psi(s)} = \frac{C_\psi}{s^2 \log^2\br{s/\Lambda^2}}.
	\label{eq.FormfactorPsi}
}
For the constant $C_\psi$ here we use the same value as it was onbtained
in the case of proton--antiproton
production \cite{Bystritskiy:2021frx} since gluons to not feel the flavour
of the quarks in the final baryons:
\eq{
	C_\psi = \br{45 \pm 9}~\GeV^4.
	\label{eq.CPsi}
}

\section{Numerical results}
\label{sec.Numerical}

\begin{figure}
	\vspace{5mm}
	\centering
    \includegraphics[width=0.45\textwidth]{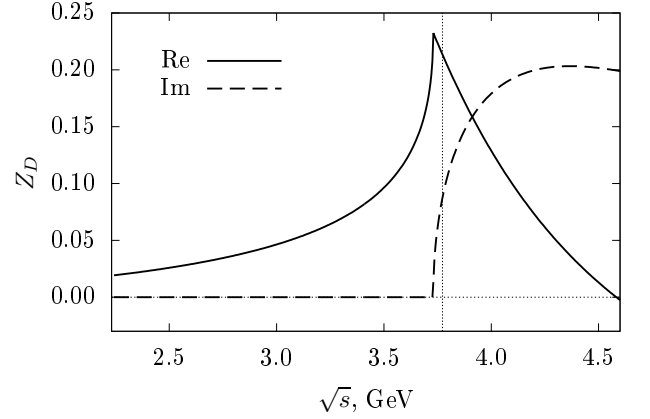}
    \caption{
    	The quantity $Z_D\br{s}$ from (\ref{eq.ZD}) as a function of
    	$\sqrt{s}$ starting from
    	the threshold $\sqrt{s} = 2M_\Lambda$.
    	The vertical dashed line shows the position of $\psi(3770)$.
    }
    \label{fig.ZD}
\end{figure}

\begin{figure}
	\vspace{5mm}
	\centering
    \includegraphics[width=0.45\textwidth]{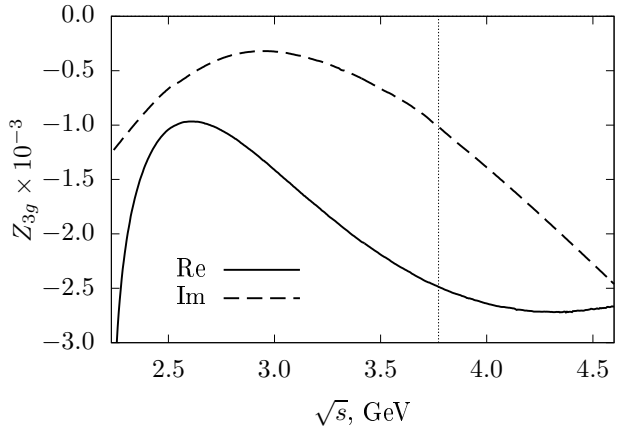}
    \caption{
    	The quantity $Z_{3g}\br{s}$ from (\ref{eq.Z3g}) as a function of
    	$\sqrt{s}$ starting from
    	the threshold $\sqrt{s} = 2M_\Lambda$.
    	The vertical dashed line shows the position of $\psi(3770)$.
    }
    \label{fig.Z3g}
\end{figure}

The main building blocks for the cross section are the quantities $Z_D\br{s}$ from (\ref{eq.ZD})
and $Z_{3g}\br{s}$ from (\ref{eq.Z3g}) which give the corresponding ($D$-meson loop and three gluon)
contributions.
In Fig.~\ref{fig.ZD} we present the dependence of $Z_D\br{s}$ as a function of total
energy $\sqrt{s}$ in the range starting from the
threshold of the reaction $\sqrt{s} = 2M_\Lambda$ up to $4.5~\GeV$. One can see that
the shape and the numerical values of the real and the imaginary parts of this quantity remain
the same as it were in the case of proton--antiproton final state (see Fig.~7 (a) in \cite{Bystritskiy:2021frx}).
As for the Fig.~\ref{fig.Z3g} presenting real and imaginary parts of the quantity $Z_{3g}\br{s}$
it shows the similar general behavior of the curves as it were in the case of proton--antiproton
final state (see Fig.~7 (b) in \cite{Bystritskiy:2021frx}) but the numerical difference is much more
seeable. Nevertheless the characteristic large negative values of this quantity still remains thus giving large
relative phase with respect to the Born contribution in the amplitude.

\begin{figure}
	\centering
    \includegraphics[width=0.48\textwidth]{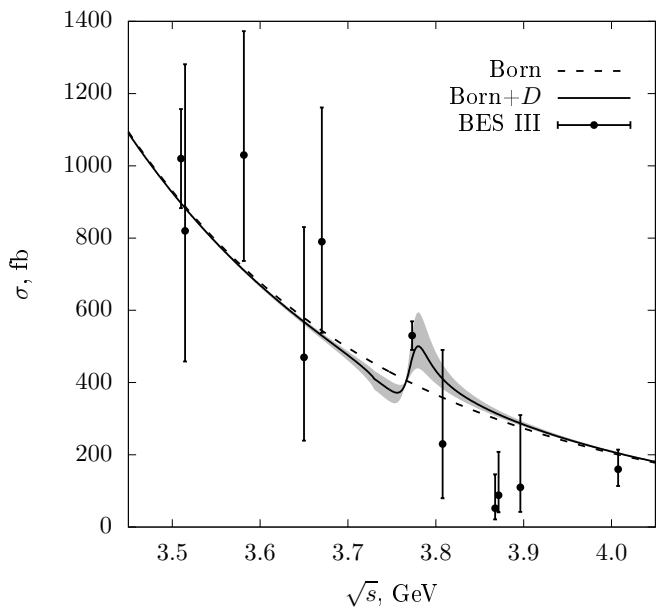}
    \caption{
    	The $D$-meson loop contributing to the total cross section with respect to the
    	BES III data \cite{BESIII:2021ccp}.
    }
    \label{fig.NarrowDLoop}
\end{figure}

\begin{figure}
	\centering
    \includegraphics[width=0.48\textwidth]{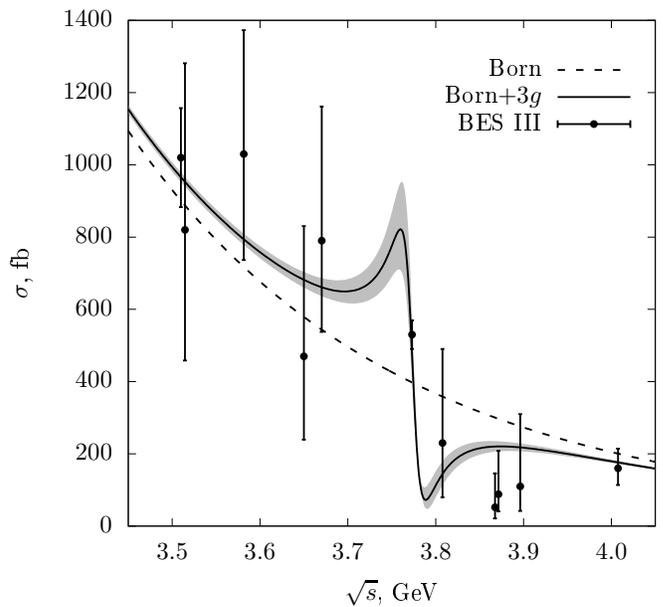}
    \caption{
    	The three gluon contributing to the total cross section with respect to the
    	BES III data \cite{BESIII:2021ccp}.
    }
    \label{fig.Narrow3g}
\end{figure}

\begin{figure}
	\centering
    \includegraphics[width=0.48\textwidth]{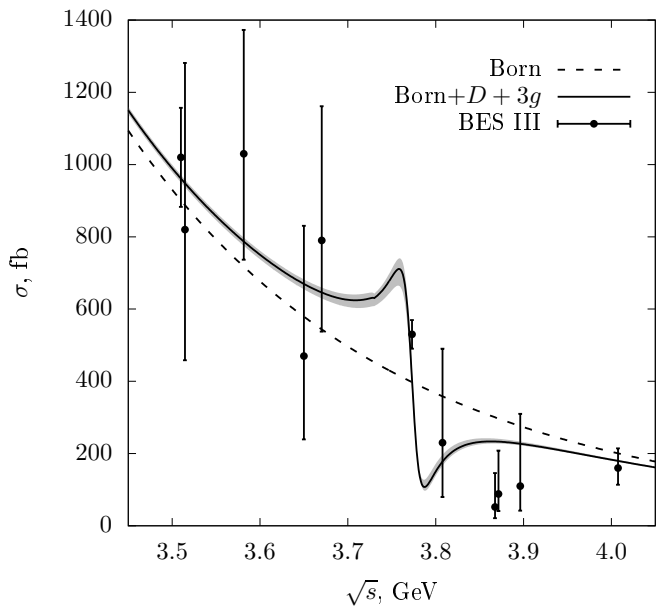}
    \caption{
    	The total cross section including two mechanisms in comparison with the
    	BES III data \cite{BESIII:2021ccp}.
    }
    \label{fig.CSNarrow}
\end{figure}

In Fig.~\ref{fig.NarrowDLoop} one can see the contributions from pure $D$-meson loop while
in Fig.~\ref{fig.Narrow3g} the pure three gluons contribution is present.
Both of these contributions are compared with the data of BES III collaboration \cite{BESIII:2021ccp}
in the vicinity of $\psi(3770)$ resonance.
For $D$-meson loop contributions, Fig.~\ref{fig.NarrowDLoop}, the error bands are provided by the errors of parameter
$g_{\Lambda D \Xi}$ from (\ref{eq.gLambdaDXi}). One can see that the central value of the curve goes lower then the experimental
point while the upper error band touches it.
For the three gluons contribution, Fig.~\ref{fig.Narrow3g}, the error bands represent
the uncertainty of the parameter $C_\psi$ from (\ref{eq.CPsi}).
We do not include to this error bands the possible uncertainties due to
the strong dependence of this mechanism of parameter $\alpha_s$ (see text after eq.~(\ref{eq.phi})).
Here we see a good agreement of this pure three gluon mechanism with the data point at $\sqrt{s} = M_\psi$.

In Fig.~\ref{fig.CSNarrow} we present the total cross section including both of these
mechanisms in comparison with the BES III data \cite{BESIII:2021ccp}.
We see rather fair agreement of our calculation with the data: the point at $\sqrt{s} = M_\psi$
is rather close to the curve and the left and right shoulders of the curve catch the
tendency of the data.
Here we must remind that we do not do any extra fit of parameters.
All the parameters of our model are fixed by the calculation for case of the proton--antiproton
final state \cite{Bystritskiy:2021frx}.

\begin{figure}
	\centering
    \includegraphics[width=0.48\textwidth]{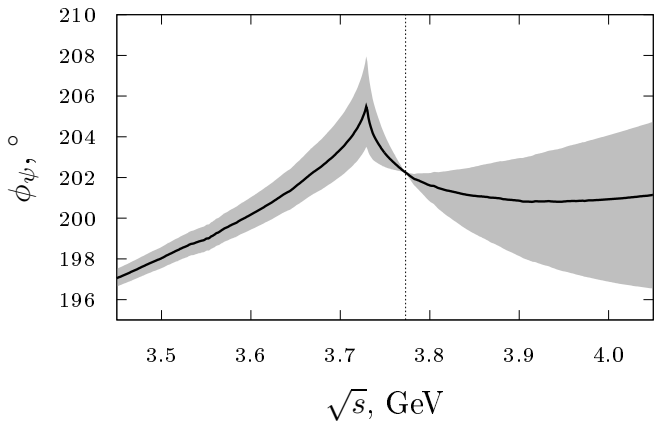}
    \caption{
    	The relative phase of total $\GeV$ charmonium $\psi(3770)$ contribution
    	as a function of $\sqrt{s}$.
    }
    \label{fig.PhiNarrow}
\end{figure}

In Fig.~\ref{fig.PhiNarrow} one can see the total relative phase $\phi_\psi$
of the charmonium contribution $\M_\psi$ to the amplitude with respect
to the Born contribution $\M_B$ without Breit-Wigner factor, i.e.:
\eq{
    S_D \br{s} + S_{3g}\br{s}
    =
    \brm{S\br{s}} e^{i \phi_\psi},
}
where $S_D \br{s}$ was defined in (\ref{eq.SD}) and
$S_{3g}\br{s}$ is from (\ref{eq.S3g}).
The error bands on this plot are due to both $g_{\Lambda D \Xi}$
and $C_\psi$ parameters uncertainties.
As one can see at the point of $\psi(3770)$ charmonium
the relative phase is rather large:
\eq{
	\phi_\psi \approx 202^{\circ}.
}
It seems that this feature is common for the charmonium
decay into two baryons final state.
We showed this in proton--antiproton final state
for the charmonium $\psi(3770)$ in papers \cite{Ahmadov:2013ova,Bystritskiy:2021frx}
and for the charmonium $\chi_{c2}(1P)(3556)$ in paper \cite{Kuraev:2013swa}.

\section{Conclusion}
\label{sec.Conclusion}

We considered the process of the electron--positron annihilation into
a $\Lambda\bar{\Lambda}$-pair in the vicinity of charmonium $\psi(3770)$ resonance.
Besides the Born mechanism, which is the pure QED, there are two contributions
related with the intermediate charmonium $\psi(3770)$ state.
One of them is the $D$-meson loop and the other is the three gluon mechanism.

It was shown that both mechanisms contribute noticeably and give much of the final result.
The total sum of them gives rather good agreement with the experimental point at
$\sqrt{s} = M_\psi$. It is also important that our curve reproduce
the tendency of the experimental points at left and at right shoulders with respect to
the central point.
It is worth to notice again that we do not use any fitting procedure in this calculation.
The parameters were fixed for the proton-antiproton production channel in the paper \cite{Bystritskiy:2021frx}.

It is extremely desirable to make a precise scan over the energy region around charmonium
$\psi(3770)$ resonance with the small steps. This could support the conclusion
that in charmonium decay the phase of $\psi \to p\bar{p}$ and $\psi \to \Lambda\bar{\Lambda}$
vertexes are large ($\phi_\psi \sim 200^\circ$) and can be precisely measured in this channels.
We showed this large phase generation in a set of papers \cite{Ahmadov:2013ova,Kuraev:2013swa,Bystritskiy:2021frx}.

In the future we plan to consider another binary final states production processes
induced by the charmonium annihilation. Polarized effects are also can be considered
since the formalism is ready and the data are present \cite{BESIII:2021cvv}.

\begin{acknowledgments}

The authors wish to express their gratitude
to Dr.~V.A.~Zykunov and Dr.~E.~Tomasi-Gustafsson for intensive and valuable discussions.

\end{acknowledgments}


\end{document}